\documentclass[twocolumn]{aastex61}
\usepackage{amssymb, amsmath}
\usepackage{graphicx}
\usepackage{natbib}
\usepackage{latexsym}
\usepackage{mathrsfs}
\usepackage{mathtools}

\begin{document}

\title{Ground-based optical transmission spectroscopy \\of the small, rocky exoplanet GJ 1132b}

\correspondingauthor{Hannah Diamond-Lowe}
\email{hdiamondlowe@cfa.harvard.edu}

\author{Hannah Diamond-Lowe}
\affil{Harvard-Smithsonian Center for Astrophysics, 60 Garden St., Cambridge, MA 02138, USA}

\author{Zachory Berta-Thompson}
\affil{Department of Astrophysical and Planetary Sciences, University of Colorado, 2000 Colorado Ave., Boulder, CO 80305, USA}

\author{David Charbonneau}
\affil{Harvard-Smithsonian Center for Astrophysics, 60 Garden St., Cambridge, MA 02138, USA}

\author{Eliza M.-R. Kempton}
\affil{Department of Physics, Grinnell College, 1116 8th Avenue, Grinnell, IA 50112, USA}
\affil{Department of Astronomy, University of Maryland, College Park, MD 20742, USA}

\begin{abstract}

Terrestrial Solar System planets either have high mean molecular weight atmospheres, as with Venus, Mars, and Earth, or no atmosphere at all, as with Mercury. We do not have sufficient observational information to know if this is typical of terrestrial planets or a phenomenon unique to the Solar System. The bulk of atmospheric exoplanet studies have focused on hot Jupiters and Neptunes, but recent discoveries of small, rocky exoplanets transiting small, nearby stars provide targets that are amenable to atmospheric study. GJ 1132b has a radius of 1.2 $R_{\oplus}$ and a mass of 1.6 $M_{\oplus}$, and orbits an M-dwarf 12 parsecs away from the Solar System. We present results from five transits of GJ 1132b taken with the Magellan Clay Telescope and the LDSS3C multi-object spectrograph. We jointly fit our five data sets when determining the best-fit transit parameters both for the white light curve and wavelength-binned light curves. We bin the light curves into 20 nm wavelength bands to construct the transmission spectrum. Our results disfavor a clear, 10$\times$ solar metallicity atmosphere at 3.7$\sigma$ confidence and a 10\% H$_2$O, 90\% H$_2$ atmosphere at 3.5$\sigma$ confidence. Our data are consistent with a featureless spectrum, implying that GJ 1132b has a high mean molecular weight atmosphere or no atmosphere at all, though we do not account for the possible presence of aerosols. This result is in agreement with theoretical work which suggests that a planet of GJ 1132b's mass and insolation should not be able to retain a H$_2$ envelope. \\

\end{abstract}

\keywords{planets and satellites: atmospheres -- planets and satellites: terrestrial planets -- planets and satellites: individual: GJ 1132b}

%%%%%%%%%%%%%%%%%%%%%%%%%%%%%%%%%%%%%%%%%%%%%%%%%%%%%%%%%%%%%%%%%%%%%%%%%%%%%%%%%%%%%%%%%%%%%%%%%%%%%%%%%%%%%%%%%%%%%%%%%%%%%%%%%%%%%%%
%%%%%%%%%%%%%%%%%%%%%%%%%% Introduction %%%%%%%%%%%%%%%%%%%%%%%%%%%%%%%%%%%%%%%%%%%%%%%%%%%%%%%%%%%%%%%%%%%%%%%%%%%%%%%%%%%%%%%%%%%%%%%
%%%%%%%%%%%%%%%%%%%%%%%%%%%%%%%%%%%%%%%%%%%%%%%%%%%%%%%%%%%%%%%%%%%%%%%%%%%%%%%%%%%%%%%%%%%%%%%%%%%%%%%%%%%%%%%%%%%%%%%%%%%%%%%%%%%%%%%

\section{Introduction} \label{sec:intro}

\indent Four years of transit data from the \textit{Kepler} mission showed us that terrestrial planets are common around low mass stars \citep{Dressing2013,Dressing2015,Gaidos2016}. The \textit{Kepler} data set also led to theories suggesting that some small planets retain hydrogen and helium envelopes from formation, comprising a small fraction of their total masses \citep{Wolfgang&Lopez2015}. These H/He envelopes are subsequently sculpted by incident extreme ultra-violet (EUV) and X-ray radiation from the host stars which, in the absence of a strong planetary magnetic field, drives atmospheric escape \citep{Ehrenreich2015}.\\
\indent M dwarfs have extended pre-main sequence phases \citep{Baraffe2015} and remain chromospherically active on long timescales \citep{Newton2017}, so it is possible that terrestrial planets orbiting M dwarfs have been stripped of any primordial atmospheres early on \citep{Lopez2013,Luger2015}. For instance, the terrestrial planets TRAPPIST-1b and c orbiting an ultracool dwarf do not exhibit transmission spectra consistent with a cloud-free low mean molecular weight atmosphere at the level of $\geq 10 \sigma$ confidence \citep{deWit2016}. TRAPPIST-1d, e, and f also do not exhibit evidence for such atmospheres at the level of $\geq 4\sigma$ confidence \citep{deWit2018}. We might expect a similar result for other small planets in close-orbits around cool stars.\\
\indent In this work we use ground-based observations to investigate the idea that terrestrial exoplanets orbiting M dwarfs do not possess low mean molecular weight atmospheres. We focus on the terrestrial exoplanet GJ 1132b (1.2 $R_{\oplus}$, 1.6 $M_{\oplus}$) orbiting a M4.5V dwarf which is 12 parsecs away from the Solar System. The radius and mass of GJ 1132b are consistent with an iron and silicate composition similar to that of Earth and Venus \citep{Berta-Thompson2015}.\\
\indent The surface gravity and estimated atmospheric temperature of GJ 1132b mean that a solar composition, hydrogen-dominated atmosphere might be detectable with ground-based instrumentation. Though we are looking for the signature of a low mean-molecular weight atmosphere, hydrogen itself is not a strong absorber, making it a difficult to detect via transmission spectroscopy. Instead, we assume the atmosphere to be well-mixed and search for tracer molecules like water (H$_2$O) or methane (CH$_4$), which have large absorption cross sections in the visible to near-infrared wavelengths.\\
\indent Understanding the nature of terrestrial exoplanet atmospheres will bolster efforts to constrain planet formation and atmospheric evolution, and ultimately inform our search for biosignatures on other worlds. We do not expect life as we know it to exist on the highly irradiated surface of GJ 1132b, but understanding the atmospheres of hot, rocky planets will contextualize an eventual search for life on cooler, habitable zone exoplanets.\\
\indent Though our current sample size of terrestrial exoplanets is small, it is important to understand them in the context of the well-studied Solar System inner planets. Whether a terrestrial exoplanet resembles Earth or Venus or Mercury has vast implications for its formation history and life-hosting capabilities. Still more intriguing is the chance to uncover terrestrial planets with compositions and characteristics unseen in the Solar System \citep[e.g., ][]{Morley2017}.\\
\indent In Section~\ref{sec:obs} we describe our observations of GJ 1132b in transit. In Section~\ref{sec:extract} we describe our customized data reduction pipeline and in Section~\ref{sec:detrend} we describe our customized data analysis pipeline. We present the results of this work in Section~\ref{sec:results}. We discuss the implications of ground-based investigations of terrestrial planet atmospheres in Section~\ref{sec:disc} and conclude with Section~\ref{sec:concl}.

%%%%%%%%%%%%%%%%%%%%%%%%%%%%%%%%%%%%%%%%%%%%%%%%%%%%%%%%%%%%%%%%%%%%%%%%%%%%%%%%%%%%%%%%%%%%%%%%%%%%%%%%%%%%%%%%%%%%%%%%%%%%%%%%%%%%%%%
%%%%%%%%%%%%%%%%%%%%%%%%%% Observatons %%%%%%%%%%%%%%%%%%%%%%%%%%%%%%%%%%%%%%%%%%%%%%%%%%%%%%%%%%%%%%%%%%%%%%%%%%%%%%%%%%%%%%%%%%%%%%%%
%%%%%%%%%%%%%%%%%%%%%%%%%%%%%%%%%%%%%%%%%%%%%%%%%%%%%%%%%%%%%%%%%%%%%%%%%%%%%%%%%%%%%%%%%%%%%%%%%%%%%%%%%%%%%%%%%%%%%%%%%%%%%%%%%%%%%%%

\begin{table*}[ht]
\centering
\caption{Observations\label{tab:obs}}
\begin{tabular}{cccccccC}
\tablewidth{0pt}
\hline
\hline
Data Set & Date & Exposure Time & Number of & & Airmass & & \mathrm{Seeing}\tablenotemark{a}\\
\cline{5-7}
  No.      & [UTC] & [s]           & Exposures & Start & Middle & End        & \mathrm{[arcsec]} \\
\hline
1 & 2016-02-28 06:01:14 -- 2016-02-28 09:15:13 & 12 & 401 & 1.109 & 1.321 & 1.849 & 0.54 \\
2 & 2016-03-04 02:28:11 -- 2016-03-04 06:29:56 & 13 & 481 & 1.119 & 1.055 & 1.190 & 0.90 \\
3 & 2016-03-08 23:50:48 -- 2016-03-09 05:41:20 & 13 & 694 & 1.523 & 1.077 & 1.136 & 0.70-1.10 \\
*  & 2016-03-21 -- 2016-03-22                  & $-$ & $--$ & $--$ & $--$ & $--$  & $---$ \\
4 & 2016-04-17 02:20:47 -- 2016-04-17 06:12:37 & 13 & 464 & 1.080 & 1.294 & 1.938 & 0.80 \\
5 & 2016-04-21 23:30:33 -- 2016-04-22 05:34:25 & 13 & 725 & 1.100 & 1.113 & 1.780 & 0.60-1.01\\
*  & 2016-05-04 -- 2016-05-05                  & $-$ & $--$ & $--$ & $--$ & $--$  & $---$ \\
*  & 2016-05-22 -- 2016-05-23                  & $-$ & $--$ & $--$ & $--$ & $--$  & $---$ \\
\hline
\end{tabular}
\begin{minipage}[t]{0.89\linewidth}
\hfill\break
\\{* We were not able to take data on these nights due to poor weather conditions.}
\tablenotetext{a}{On nights 1, 2, and 4 the seeing remained relatively stable throughout the night while on nights 3 and 5 the seeing deteriorated over the course of the observations. }
\end{minipage}
\end{table*}

\section{Observations} \label{sec:obs}

A joint program between Harvard and MIT (PIs Diamond-Lowe and Berta-Thompson, respectively) to observe transits of GJ 1132b received eight nights on the Magellan II (Clay) Telescope with the LDSS3C\footnote{\url{www.lco.cl/telescopes-information/magellan/instruments/ldss-3}} multi-object spectrograph at Las Campanas Observatory \citep{Stevenson2016a}. Of the eight observing opportunities we observed five transits of GJ 1132b and lost the remaining three nights to clouds and high winds. The details of our observing program are presented in Table~\ref{tab:obs}.\\
\indent GJ 1132 (V = 13.49, K = 8.322) is an M4.5V star \citep{Berta-Thompson2015}. In the 4$'$ field of view of LDSS3C there are no stars of comparable magnitude or spectral type, so we opted to simultaneously observe nine comparison stars which we later used to remove telluric effects from the GJ 1132 spectrum. Of these comparison stars one was brighter than GJ 1132 but it saturated our detector and we were not able to use it in our analysis.\\
\indent Our LDSS3C masks include slits for GJ 1132 and the nine comparison stars. At the time of our observations there was a background star 7.3 arcseconds away form GJ 1132; because GJ 1132 is a high proper motion star this separation will change over time and future observers of GJ 1132 should account for this. We oriented our mask such that the background star did not contaminate the dispersed spectrum of GJ 1132. We cut our slits 10$''$ in width to avoid slit losses and 20$''$ in length to provide sky background with which to perform our subtraction \citep{Bean2010}. We also cut identical masks with 1$''$ wide slits which we used to take wavelength calibration arcs during the afternoon prior to each observation. \\
\indent We set the detector binning to 2x2 and the readout speed to \texttt{Fast} (the LDSS3C user manual says this will give a 13 second read out time but we found it to be 16 seconds). We set the gain to \texttt{Low}, which, along with the readout speed, gives a gain of 0.6 ADU/electron. With our observation mask we took biases, darks, quartz flat fields, and a mask image with which to align our stars in the slits during observations. With our 1$''$ mask we took helium, neon, and argon arcs so that we could determine a wavelength solution for each dispersed stellar spectrum. Both during calibrations and observations we kept every detector pixel that we used to perform our analysis below 53,000 ADU. As stated in the LDSS3C user manual and corroborated by the Las Campanas Observatory instrument specialists, the full pixel well is 65,536 ADU, but past 53,000 ADU the detector stops counting photoelectrons linearly. \\
\indent We chose to use the VPH-Red grism which provides a wavelength coverage of 640-1040 nm with a central wavelength of 850 nm and a linear dispersion of 0.1175 nm$/$pixel \citep{Stevenson2016a}. The VPH-Red grism has a higher resolution than the VPH-all grism, as well as a higher throughput at redder wavelengths. Using the VPH-Red grism allowed us to take longer exposures without saturating the detector, while also focusing on those wavelengths where GJ 1132 is brightest.\\
\indent We took 13 second integrations and achieved a duty cycle of 45\%. The VPH-Red grism introduces order contamination onto the detector, which we mitigated with the OG590 order-blocking filter as advised in the LDSS3C user manual. This filter blocks spectral contamination from higher spectral orders but produces internal reflections. (\citet{Stevenson2016a} noted this contamination and decided against using the OG590 filter.) After inspecting the calibration arc frames during the day we decided that the OG590 contamination was less problematic than the higher-order line contamination. We therefore used the OG590 filter during observation and also while taking our calibration images. \\
\indent We note that our first night of observation (data set number 1 in Table~\ref{tab:obs}) differed from the rest for two reasons. Firstly, we neglected to use the OG590 order blocking filter, which is why we exposed for 12 seconds on this night instead of 13. In spite of this, the order contamination was not drastic since GJ 1132 emits few photons blue-ward of 700 nm. Secondly, we used a slightly different mask. The first amplifier (C1) of LDSS3C's CCD has several columns of bad pixels which over-lapped with one of our comparison stars. We cut a second, identical mask with the slits slightly shifted in order to avoid the bad pixels. We did not end up using this comparison star since the bad pixels near it saturated and leaked light into its dispersed spectrum. For consistency we exclude this comparison star from all five data sets when performing the analysis. \\
\indent For all five of our data sets we acquired at least one transit-durations's worth of out-of-transit baseline both before and after the transit event with which to estimate the basline flux and correct for correlated noise in the data.

%%%%%%%%%%%%%%%%%%%%%%%%%%%%%%%%%%%%%%%%%%%%%%%%%%%%%%%%%%%%%%%%%%%%%%%%%%%%%%%%%%%%%%%%%%%%%%%%%%%%%%%
%%%%%%%%%%%%%%%%%%%%%%%%%%%%%%%%%%%%% Data Extraction %%%%%%%%%%%%%%%%%%%%%%%%%%%%%%%%%%%%%%%%%%%%%%%%%%
%%%%%%%%%%%%%%%%%%%%%%%%%%%%%%%%%%%%%%%%%%%%%%%%%%%%%%%%%%%%%%%%%%%%%%%%%%%%%%%%%%%%%%%%%%%%%%%%%%%%%%%%

\section{Data Extraction} \label{sec:extract}

\begin{figure}
\includegraphics[scale=.33]{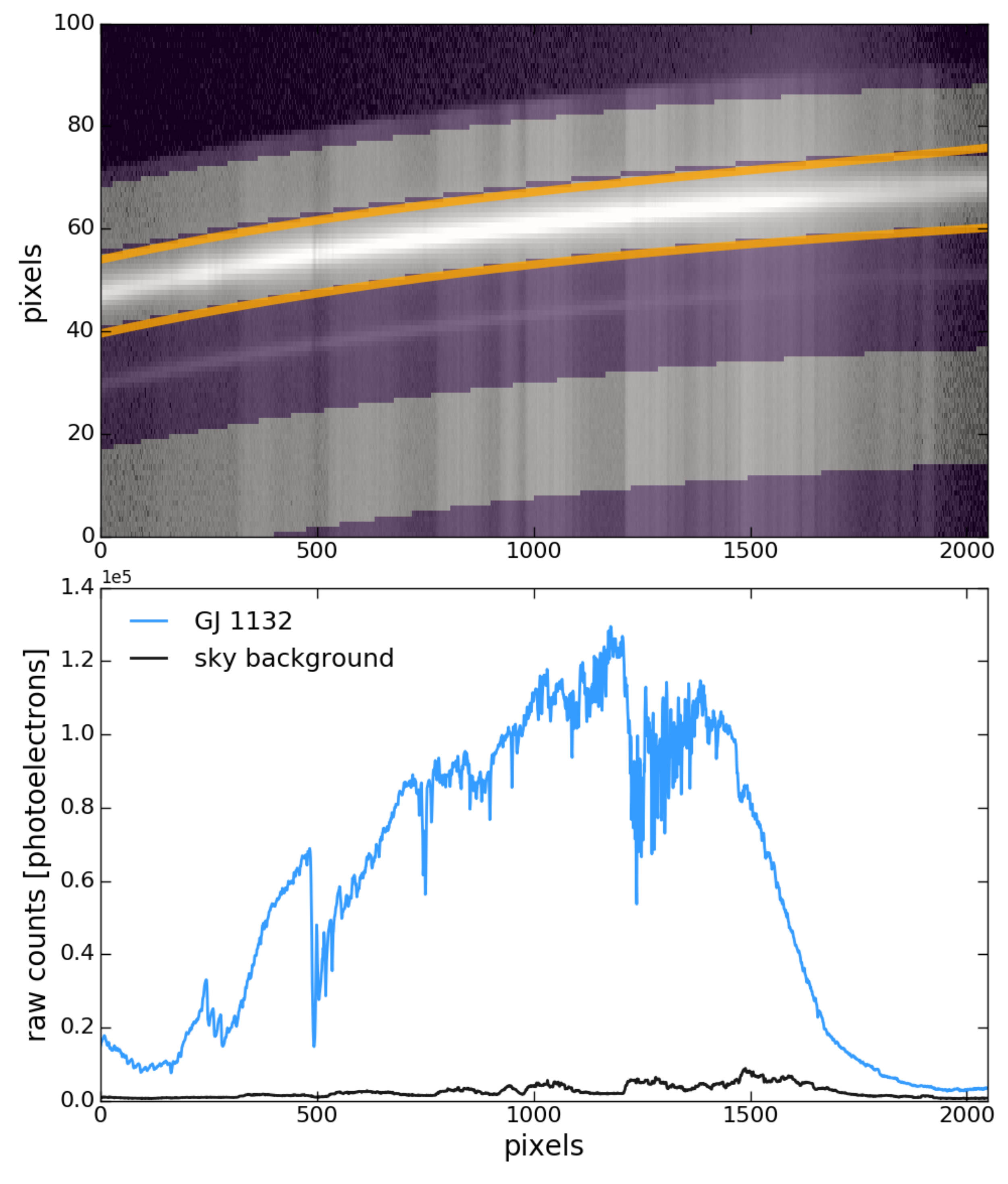}
\caption{Intermediary steps in the \texttt{mosasaurus} open source extraction pipeline for multi-object spectrographs. This figure corresponds to a single spectrum of GJ 1132. \textit{Top}: Spectral trace of GJ 1132 in which the curvature is apparent. Orange lines show the bounds of the extraction aperture. Shaded purple regions are data that we discard when doing our analysis. This includes a region directly beneath the GJ 1132 trace that is masking out the spectrum of a faint background star.\textit{Bottom}: Extracted 1D raw spectrum of GJ 1132 prior to wavelength calibration (light blue line). Also shown is the 1D sky background spectrum which is removed from the GJ 1132 spectrum (black line)}\label{fig:ext}
\end{figure}

\indent We transform our raw Magellan/LDSS3C images into 1D stellar spectra by running them through our custom Python pipeline, \texttt{mosasaurus}\footnote{\url{github.com/zkbt/mosasaurus}, v0.0}. With this pipeline, we perform basic CCD processing on pairs of FITS images from the two amplifiers on LDSS3C. After subtracting 1D biases estimated from the amplifiers' overscan regions, we stitch images together into full frames, using the amplifiers' reported gains to convert from ADU to electrons. We create median-stacked 2D bias and dark exposures that we subtract from all quartz flat and science exposures, to remove the baseline level of the readout electronics and the (very small) dark current accumulated during all exposures.\\
\indent To identify and mitigate cosmic ray contamination, we compare each image to the 10 closest images in time. For each pixel, we calculate the median absolute deviation from the median (MAD), and flag any upward outliers that exceed $10\times$ MAD as cosmic ray hits. We replace the flux value in the pixel affected by cosmic rays with its median value from the immediately surrounding exposures. We keep track of which pixels have been modified in this fashion, so they can be masked out of later analysis stages if so desired. \\
\indent We cut out a 60$\times$2048 pixel region around each of our dispersed spectra. We cut out corresponding regions using the same pixels on our quartz flat and arc images. The spectra recorded on the detector are curved slightly (by about 10 pixels over the entire thousand-pixel chip). We fit a second order polynomial to the spectral trace, which maps where the peak flux is in each column in the cross-dispersion direction. We calculate the full-width-half-maximum (FWHM) in each column. To create an extraction window we extend by three times the FWHM from the centroid in the spatial direction (Figure~\ref{fig:ext}), allowing the extraction width to vary with wavelength. We create a range of extraction window sizes for each stellar spectrum, for later comparison. These apertures remain fixed with respect to the detector; they do not move to follow the slight motion of the spectral trace throughout the night (1-3 pixels). \\
\indent We use an interactive tool to plot the science images with an overlayed extraction window, to inspect the aligned extraction window containing the stellar flux, and to set custom sky subtraction regions uniquely for each star (Figure~\ref{fig:ext}). For the GJ 1132 field, we determine at this stage that several comparison stars are unusable -- the bright one that saturated the detector and four others that turned out to have multiple stars clustered together in the slit. Having multiple stars in a single slit is problematic as we would have to combine their spectra in a large extraction window, which leads to a poor estimate of the sky background and the blending of different wavelengths from different stars in the same extracted spectrum. We end up with four comparison stars for our analysis. Though we use the same comparison stars for each night of data the extraction windows may vary from night to night. This is because the seeing conditions on a given night influence the PSF of the stars on the detector. We therefore stand to benefit from using different extraction windows for each star for each data set. \\
\indent We median-pass filter quartz flat exposures, taken through the same wide slits as our science data, by dividing each pixel by the median of the 20$\times$100 pixels surrounding it. We then divide each spectrum region in the time-series by this filtered quartz flat to correct for the intrinsic pixel-to-pixel inconsistencies of the detector. We create a 1D stellar spectrum from each flat-fielded stellar region by summing up the flux in each column in the spatial direction, accounting for partial pixels at the edges of the extraction aperture. We create a 1D sky-background spectrum from each flat-fielded stellar region by fitting a two-degree polynomial to each column in the spatial direction outside the extraction window, and then summing over the column. We then subtract the sky-background spectrum from the stellar spectrum.\\
\indent We tested an optimal extraction routine as outlined by \citet{Horne1986}. We find that this method makes at most a 10\% improvement in signal-to-noise for the faintest comparision stars, but does not improve signal-to-noise for GJ 1132 or the brighter comparision stars. Because the fainter comparison stars have a proportionately small influence on the resulting light curve, we use the extraction method outlined above and not the optimal extraction routine.\\
We use the He, Ne, and Ar arcs taken during calibration to develop a rough wavelength solution for each star. The LDSS3C user manual provides a wavelength solution for the VPH-Red grism that gives the pixel position of prominent features in the He, Ne, and Ar spectra. Using a customized graphical user interface we match up the features in the provided wavelength solution to those in each arc, corresponding to our stellar spectrum regions, and create a polynomial wavelength solution for each star. In practice, this works better for some stars than others, but it generally lines up the spectra with each other to within 5 nm.\\
\indent We then choose one exposure of one star as a basis against which to cross-correlate all of the exposures of all the stars in a given data set. We use five prominent features in the spectra in order to perform the cross-correlation: the O$_2$ doublet (760.5 nm), each line of the Ca triplet (849.8, 854.2, and 866.2 nm), and the forest of water lines (about 930-980 nm). We note that the Ca triplet is not a telluric feature and so may be risky to use when calibrating the spectra. In this case, all of the stars we observe are in the Sun's local moving group, and any Doppler shifting of the Ca lines are not detectable at the velocity dispersion of the LDSS3C spectrograph and the VPH-Red grism (about 165 km/s/pixel). Given the small field-of-view of the instrument we are not concerned about the different lines-of-sight to each star.\\ \indent This process reveals that there is both a shifting and stretching of the spectra over the course of the observations. For instance, in data set number 1, the difference between the positions of the O$_2$ doublet and the water line forest in the GJ 1132 spectrum increases by a pixel from the start of the observation relative to the end. We use this information to apply a second wavelength solution for each spectrum in each exposure such that they are lined-up with one another in wavelength space to within 0.35 nm across all stars and the entire night. This is the final step in achieving 1D spectra which we can use to make our light curves.

%%%%%%%%%%%%%%%%%%%%%%%%%%%%%%%%%%%%%%%%%%%%%%%%%%%%%%%%%%%%%%%%%%%%%%%%%%%%%%%%%%%%%%%%%%%%%%%%%%%%%%%%%%%%%%%%%%%%%%%%%%%%%%%%%%%%%
%%%%%%%%%%%%%%%%%%%%%%%%%%%%%%%%%%%%%% Data Analysis %%%%%%%%%%%%%%%%%%%%%%%%%%%%%%%%%%%%%%%%%%%%%%%%%%%%%%%%%%%%%%%%%%%%%%%%%%%%%%%
%%%%%%%%%%%%%%%%%%%%%%%%%%%%%%%%%%%%%%%%%%%%%%%%%%%%%%%%%%%%%%%%%%%%%%%%%%%%%%%%%%%%%%%%%%%%%%%%%%%%%%%%%%%%%%%%%%%%%%%%%%%%%%%%%%%

\section{Data Analysis} \label{sec:detrend}

To perform this analysis we constructed the code \texttt{detrendersaurus}\footnote{github.com/hdiamondlowe/detrendersaurus, v1.0}. Though it is not generalized for data sets other than LDSS3C multi-object spectroscopy, the code is fairly modular and some routines may be useful to others performing similar analyses.\\

\subsection{Analyzing transits separately} \label{subsec:separate}

\begin{figure}
\includegraphics[scale=.42]{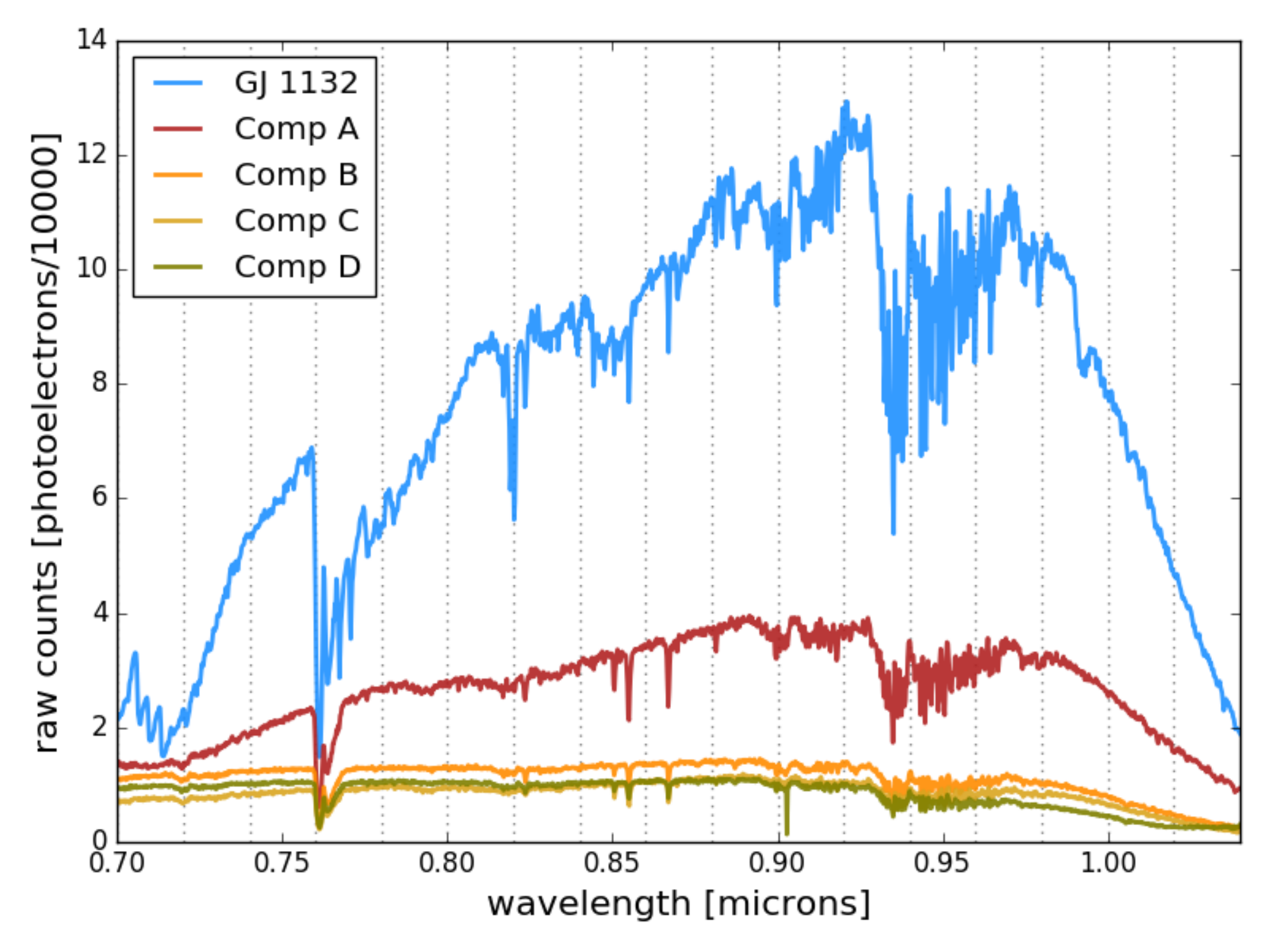}
\caption{Wavelength-calibrated spectra of GJ 1132 and the four stars we use to remove telluric features from the GJ 1132 spectrum. The vertical dotted lines show the boundaries of the wavelength we use to make our transmission spectrum. The comparison stars are all fainter than GJ 1132. By summing the comparison stars' flux we achieve 71\% of GJ 1132's flux when integrating over the full wavelength bandpass (700-10400 nm). This means that our results are limited by the combined photon noise of the comparison stars.} \label{fig:spectra}
\end{figure}
\begin{table}
\centering
\caption{Stars used in this work \label{tab:stars}}
\begin{tabular}{cccc}
\tablewidth{0pt}
\hline
\hline
Star & RA & Dec & Flux/\\
 & & & GJ 1132 Flux\\
\hline
GJ 1132 & 10:14:50.09  &  -47:09:17.5 & 1.0 \\
Comp A  & 10:14:57.51   &  -47:05:39.9 & 0.35  \\
Comp B  & 10:14:58.22   &  -47:09:35.1 & 0.14  \\
Comp C  & 10:15:05.74   &  -47:07:43.9 & 0.11  \\
Comp D  & 10:15:16.26   &  -47:06:44.3 & 0.11  \\
\hline
\end{tabular}
\begin{minipage}[t]{0.89\linewidth}
\hfill\break
\\{\textbf{Note.} The relative flux column indicates the full wavelength-band integrated flux of each star relative to that of GJ 1132. The comparison star labels in this table correspond to those in Figure~\ref{fig:spectra}. GJ 1132 is a high proper motion star. Positions are given for an epoch of 2016.3.}
\end{minipage}
\end{table}

\indent GJ 1132 is brighter than the four comparison stars (Figure~\ref{fig:spectra}, Table~\ref{tab:stars}). We therefore create our light curves by summing up the flux from the comparison stars and dividing the GJ 1132 spectrum by the summed comparison spectrum for each point in the light curve. GJ 1132 is still brighter than the summed flux of the four comparison stars so we are limited by the photon noise of the summed comparison star flux. \\
\indent We detrend our light curve using decorrelation parameters that either have the same values for all the stars (e.g., airmass) or are associated with GJ 1132 (e.g., width of the spectral trace). The parameters that are unique to each star have similar values for all stars in the data set but because we detect the most photons from GJ 1132 its decorrelation parameters have higher signal-to-noise ratios. We create white light curves for each data set and also bin the light curves from each data set into narrow wavelength bands for the purpose of atmospheric characterization. We restrict our analysis to the wavelength range common to GJ 1132 and the four comparison stars, which is 700-1040 nm.\\
\indent We determine which linear combination of decorrelation parameters are necessary to remove the effects of correlated noise (discussed in greater detail in Section~\ref{subsubsec:lmfit}). In a given data set we choose a single 20 nm wavelength bin without any prominent stellar, telluric, or atmospheric features (we use 830 - 850 nm) and calculate the Bayesian Information Criterion (BIC) value for every combination of possible decorrelation parameters. We check that there is no correlation with the sky-background, as this would imply that we are not properly removing the sky background during extraction. Once this process is done for all five data sets we take the union of all the best decorrelation parameters and marginalize over them in all wavelength bins in all data sets. A list of these parameters, what vectors they depend on, and how they are derived can be found in Table~\ref{tab:detrendparams}.

\begin{table*}[ht]
\centering
\caption{Decorrelation parameters used to model data systematics\label{tab:detrendparams}}
\begin{tabular}{lll}
\tablewidth{0pt}
\hline
\hline
Parameter & Vector & Description  \\
\hline
airmass & $t$ & average airmass of the field\\
rotator angle & $t$ & instrument rotator angle\\
width & $t$, star & median width across wavelengths of the stellar trace in the cross-dispersion direction\\
stretch & $t$, star & wavelength solution coefficient associated with spectrum stretching in the dispersion direction \\
peak & $t$, $\lambda$, star & brightness of the brightest pixel in the cross-dispersion direction\\
normalization & $t$ & unit array \\
\hline
\end{tabular}
\begin{minipage}[t]{0.89\linewidth}
\hfill\break
\\{\textbf{Note.} All parameters are functions of time $t$. They can also vary by wavelength $\lambda$ and by star. For all parameters that are star-dependent we use the values associated with GJ 1132 as it has the highest signal-to-noise ratio. }
\end{minipage}
\end{table*}

\indent From the results of a Levenberg-Marquardt minimizer we run a makeshift Bayesian test in order to determine whether our five nights of data should be analyzed separately or taken together in a joint fit. For each 20 nm bin in each of the five data sets we compare the resulting $\chi^2$ value for a fit in which the transit depth is allowed to vary to one in which the transit depth is fixed to a inverse-variance weighted depth derived from the five nights. We account for the change in the number of fitted parameters between these two scenarios. We find that the $\chi^2$ values for the case of the fixed transit depth in a given wavelength bin can be higher, lower, or identical to the case where the transit depth parameter is allowed to vary. In other words, fixing the transit depth does not provide a uniformly worse fit. We therefore decide to fit the five nights of data jointly, allowing the transit parameters to be shared across all nights.\\

\subsection{Analyzing transits jointly} \label{subsec:joint}

\subsubsection{Levenberg-Marquardt fits} \label{subsubsec:lmfit}

\indent In analyzing the transits jointly we must account for the different uncertainties associated with the individual data sets, as well as clip outlying data points. We use a three-step Levenberg-Marquardt process to settle on initial guesses for our parameters to use in a dynamic nested sampler, which will be discussed in further detail in Section~\ref{subsubsec:dynesty}. To run our Levenberg-Marquardt fits we employ the open-source \texttt{lmfit} package \citep{Newville2016}.\\
\indent In the first pass at the Levenberg-Marquardt fit we build a linear model unique to each night of data following the formula

\begin{equation}
\mathcal{M}(t) = \mathcal{S}(t)\mathcal{T}(t)
\end{equation}

\noindent where $\mathcal{S}(t)$ is the systematics model and $\mathcal{T}(t)$ is the transit model. The systematics model $\mathcal{S}(t)$ can further be broken down to

\begin{equation}
\mathcal{S}(t) = 1 + \sum_{n = 1}^{N} a_np_n(t)
\end{equation}

\noindent where $N$ is the number of decorrelation parameters used in the fit, $a_n$ are the coefficients we are fitting for, and $p_n$ are the arrays of decorrelation parameters that describe the correlated systematics in the data, which are all functions of time. For decorrelation parameters which are functions of wavelength we sum over wavelength space corresponding to the wavelength bin we are working in. The decorrelation parameters are either common to all stars (airmass and rotator angle) or are taken from the GJ 1132 spectral extraction (width, stretch, peak). \\
\indent We build the transit model $\mathcal{T}(t)$ using the open-source \texttt{batman} code \citep{Kreidberg2015} and feed in the free transit parameters. The transit parameters that can be shared across the five data sets are the planet-to-star radius ratio $R_p/R_*$, period $P$, inclination $i$, scaled orbital distance $a/R_*$, and uncorrelated quadratic limb darkening coefficients $2u_0 + u_1$ and $u_0 - 2u_1$ as used by \citet{Holman2006}. The residuals that we calculate from dividing our light curves by the linear models are weighted by the calculated photon noise of each data set.\\
\indent At this stage we fix the uncorrelated quadratic limb darkening coefficients to values derived from the Limb Darkening Tool Kit (\texttt{ldtk}), an open-source package that takes in stellar parameters and uncertainties and calculates the limb darkening coefficients in a given wavelength range based on the \uppercase{phoenix} library of stellar models \citep{Husser2013, Parviainen2015}. During the next stage of analysis (Section~\ref{subsubsec:dynesty}) we instead allow the uncorrelated quadratic limb darkening parameters to vary within a prior. \\
\indent In the second Levenberg-Marquardt fit we calculate the MAD of the residuals and clip the 29 data points (for the white light curve) or $\leq$ 27 points (for the wavelength-binned light curves) that deviate by 5$\times$ the MAD. In the third Levenberg-Marquardt fit we change the weighting from the calculated photon noise to the uncertainties we derive from each night's data as a result of our second fit. Levenberg-Marquardt fits with \texttt{lmfit} are inexpensive and quick but running a dynamic nested sampler can be expensive if the priors are too wide. Since we derive our sampling priors from the covariance matrix output by the Levenberg-Marquardt fit we find it expedient to constrain the fit parameters as much as possible at this stage.\\

\subsubsection{Dynamic nested sampling} \label{subsubsec:dynesty}

\indent Our joint fit comprises a minimum of 30 free parameters -- the same six decorrelation parameters (Table~\ref{tab:detrendparams}) to fit for each of the five data sets. In addition to this there can be free transit model parameters, like the transit midpoint for each night or the transit depth, which is shared between the five nights. Which transit parameters are free depends on whether or not we are performing a white light curve fit or a wavelength-depending light curve fit. With so many free parameters traditional Markov Chain Monte Carlo ensemble samplers such as \texttt{emcee} \citep{Foreman-Mackey2013} are slow and inefficient at exploring the parameter space \citep{Huijser2015}. We instead use the open-source dynamic nested sampling package \texttt{dynesty}\footnote{github.com/joshspeagle/dynesty} (J. Speagle, private communication) to estimate our posteriors. \\
\indent The \texttt{dynesty} code samples each free parameter from 0 to 1 and so requires a prior transform function to map the outputs from the sampling onto the parameter space we want to explore. For all but the uncorrelated quadratic limb darkening coefficients we set uniform priors on the parameters used to model the systematic and transit portions of our models. When possible we assume the same uniform priors for the transit model parameters as used by \citet{Dittmann2017b}. Otherwise we set uniform priors by taking the estimated 1$\sigma$ uncertainties from the covariance matrix of our Levenberg-Marquardt fit and multiplying by 25 such that the prior bounds for each parameter are 25$\sigma$ from the estimated parameter value. These wide uniform priors allow for an uninformed, broad parameter space for the sampler to explore.\\
\indent Following the work of \citet{Berta2012} we place Gaussian priors on the uncorrelated quadratic limb darkening coefficients $2u_0 + u_1$ and $u_0 - 2u_1$. To determine what these Gaussian priors should be we first get the quadratic limb darkening coefficients in each wavelength bin from \texttt{ldtk} \citep{Parviainen2015}. \texttt{ldtk} has an option to run a Markov chain Monte Carlo (MCMC) with the input stellar parameters and uncertainties in order to derive the limb darkening coefficients. We use the samples from the MCMC to calculate arrays of uncorrelated parameters using the formulation $2u_0 + u_1$ and $u_0 - 2u_1$, where $u_0$ and $u_1$ are the quadratic coefficients derived with \texttt{ldtk}. We calculate the median and standard deviation of these uncorrelated arrays and use these values to set the Gaussian priors. These Gaussian priors leverage our knowledge of stellar astrophysics without having to place complete faith in the accuracy of the stellar models.\\
\indent Also following \citet{Berta2012} we achieve a $\chi^2$ value of unity by including a rescaling parameter $s$ (Equations 2 and 3 of that paper). We automatically marginalize over this during our dynamic nested sampling by modifying our log-likelihood function such that $s$ is a multiplier of the theoretical uncertainty associated with each data point, including all terms that depend on $s$. Each data set has its own value of $s$ associated with it. An $s$ value of unity implies that we are reaching the theoretical photon noise limit with our fits, while a value less than unity implies an over-fitting of the model to the data.

\subsubsection{White light curve} \label{subsubsec:whitelc}

\begin{table}
\centering
\caption{White light curve transit model parameter priors\label{tab:params}}
\begin{tabular}{LLL}
\tablewidth{0pt}
\hline
\hline
\mathrm{Parameter} & \mathrm{Value} & \mathrm{Priors}\\
\hline
\delta t_{0,1}\ \mathrm{[days]} & -0.0015 & [-0.0075, 0.0044]\tablenotemark{a} \\
\delta t_{0,2}\ \mathrm{[days]} & -0.0017 & [-0.0057, 0.0023] \tablenotemark{a}\\
\delta t_{0,3}\ \mathrm{[days]} & -0.0019 & [-0.0108, 0.0069] \tablenotemark{a}\\
\delta t_{0,4}\ \mathrm{[days]} & -0.0016 & [-0.0093, 0.0060] \tablenotemark{a}\\
\delta t_{0,5}\ \mathrm{[days]} & -0.0017 & [-0.0060, 0.0027] \tablenotemark{a}\\
R_p/R_*    & 0.0493   & [0.0081, 0.0904]\tablenotemark{a} \\
P\ \mathrm{[days]}  & 1.628925 & [1.628744, 1.629116]\tablenotemark{b}\\
i  & 88.68 & [85, 90]\tablenotemark{b}\\
a/R_* & 16.54 & [12, 20]\tablenotemark{b}\\
2u_0 + u_1   & ---       & 0.8756 \pm 0.0128\tablenotemark{c}\\
u_0 - 2u_1  & ---       & -0.3672 \pm 0.0566\tablenotemark{c}\\
s_{1, 2, 3, 4, 5} & 1 & [0.01, 10]\tablenotemark{d}\\
\hline
\end{tabular}
\begin{minipage}[t]{0.89\linewidth}
\tablenotetext{a}{Uniform priors that are 25$\times$ the 1$\sigma$ uncertainties taken from the \texttt{lmfit} covariance matrix, as described in Section~\ref{subsubsec:dynesty}. The $\delta t_0$ parameter is the offset from the calculated time of mid-transit (Equation~\ref{eqn:t0}). $R_p/R_*$ is the planet-to-star radius ratio.}
\tablenotetext{b}{Uniform priors taken from \citet{Dittmann2017b}. $P$ is the period, $i$ is the inclination, and $a/R_*$ is the scaled orbital distance.}
\tablenotetext{c}{Gaussian priors calculated with \texttt{ldtk} outputs, as described in Section~\ref{subsubsec:dynesty}, given as mean $\pm$ standard deviation. $2u_0 + u_1$ and $u_0 - 2u_1$ are the uncorrelated quadratic limb darkening parameters. In the Levenberg-Marquardt fits these parameters are fixed to the \texttt{ldtk} outputs, but when sampling the parameter space with \texttt{dynesty} we use the Gaussian priors; \texttt{dynesty} does not require starting values as inputs.}
\tablenotetext{d}{Wide uniform priors set by hand. Each data set has a rescaling parameter $s$ as described in Section~\ref{subsubsec:dynesty}}
\end{minipage}
\end{table}

\begin{figure}
\includegraphics[scale=.425]{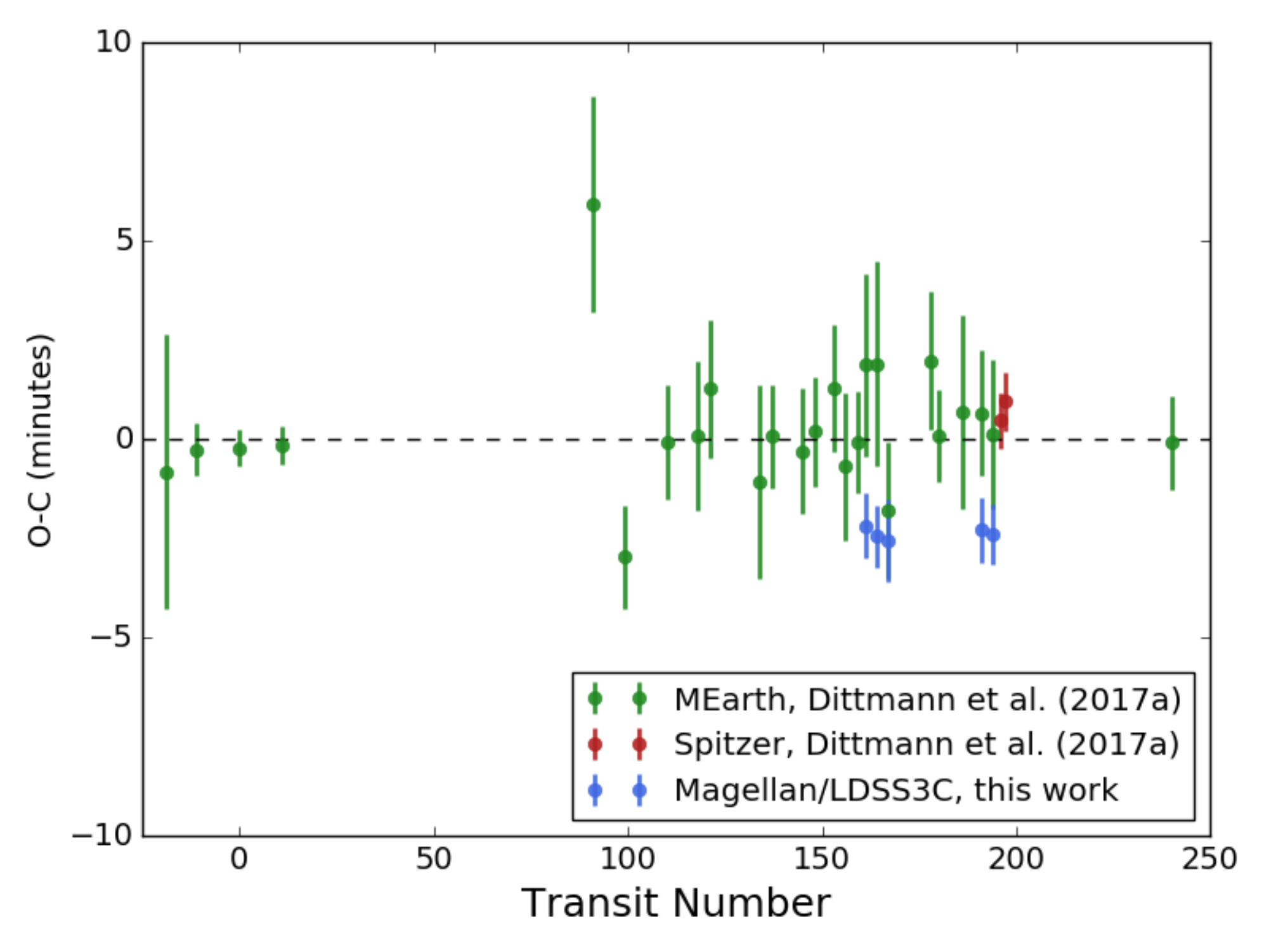}
\caption{Observed minus calculated (O-C) times of mid-transit for GJ 1132b by transit number with 1$\sigma$ error bars derived from fitting each transit. Values for MEarth (green data points) and \textit{Spitzer} (red data points) are taken from \citet[Table 4]{Dittmann2017b}. The \textit{Spitzer} points are corrected here to include leap seconds. Values for the data presented in this work from the Magellan/LDSS3C instrument are shown in blue. All values were converted to BJD$_{\mathrm{TDB}}$ for the purpose of direct comparison. We use the values of $P = 1.6289246$ days and $T_0 = 2457184.55804$ days \citep{Dittmann2017b} to claculate all times of mid-transit.} \label{fig:o-c}
\end{figure}

\begin{figure}
\includegraphics[scale=.42]{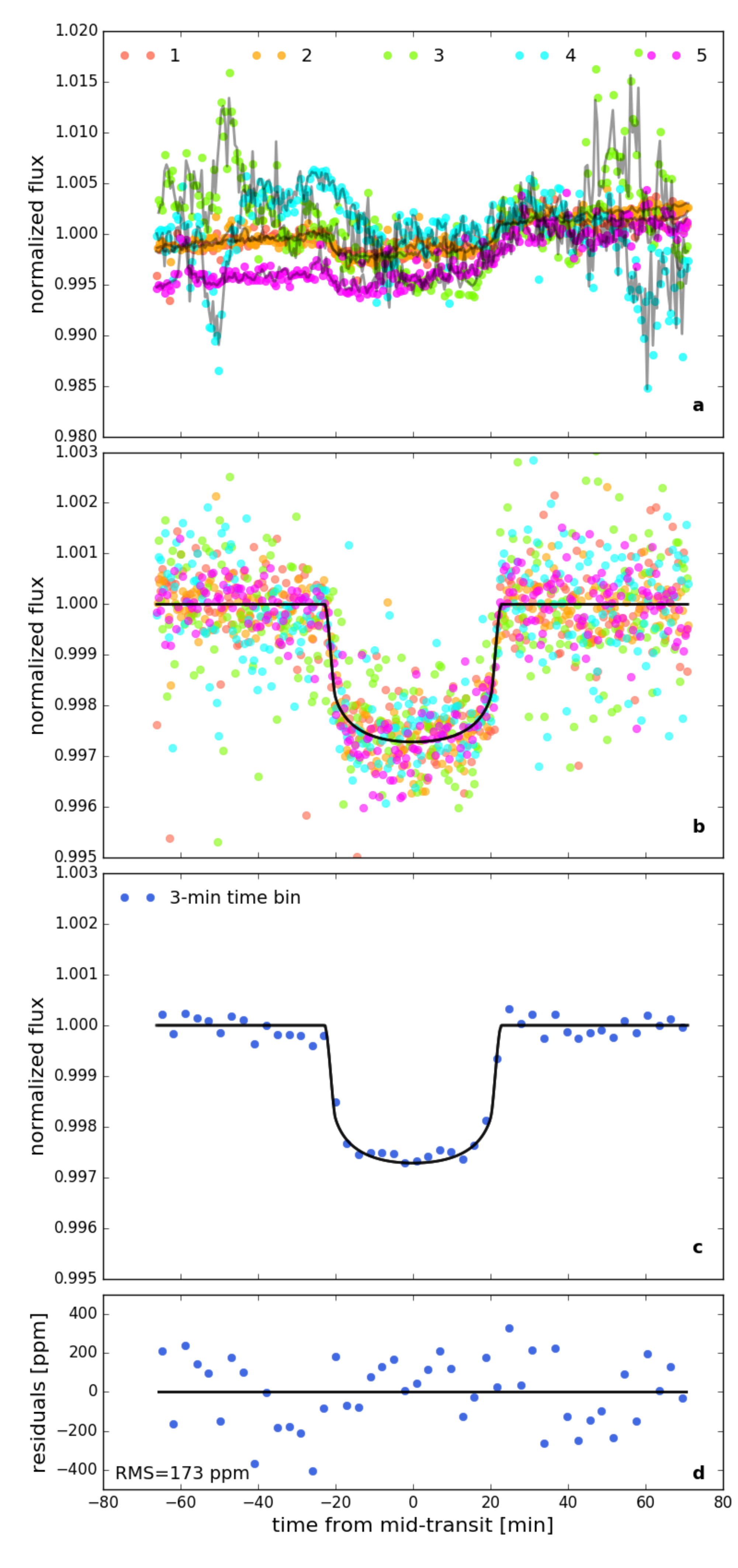}
\caption{\textit{Panel a}: Raw white light curves integrated from 700 - 920 nm from each of the five data sets with models over-plotted in grey. The systematic parameters for these models are unique to each data set but the transit parameters are free and shared jointly between the data sets. The derived values for the transit parameters are given in Table~\ref{tab:fitparams}. \textit{Panel b}: Unbinned white light curves from the five data sets with the systematics component of the models divided out. The over-plotted black line is the transit model. \textit{Panel c}: White light curve binned in time at a 3-minute cadence. The over-plotted black line is the transit model. \textit{Panel d}: Residuals after dividing the systematics models and subtracting the transit models from the raw white light curves and binning at a 3-minute cadence.} \label{fig:whitelc}
\end{figure}

\indent We jointly fit the white light curve of our five data sets and allow the time of mid-transit $\delta t_0$ to vary for each data set, along with the shared parameters of the radius ratio $R_p/R_*$, period $P$, inclination $i$, scaled orbital distance $a/R_*$, and uncorrelated quadratic limb-darkening coefficients $2u_0 + u_1$ and $u_0 - 2u_1$. In doing so we leverage the five nights of data, which have the same transit model parameter values, except for the mid-transit time.\\
\indent Where appropriate we adopt the same priors as those quoted by \citet{Dittmann2017b} (Table~\ref{tab:params}). The photometric bandpass of MEarth is not identical to the wavelength coverage of our white light curves, and so we use stellar models to set Gaussian priors on the uncorrelated quadratic limb darkening coefficients, as described in Section~\ref{subsubsec:dynesty}.\\
\indent For the time of mid-transit we fit for an offset $\delta t_0$ from the calculated mid-transit time using the ephemeris $T_0$ given by \citet{Dittmann2017b}:
\begin{equation}\label{eqn:t0}
\delta t_0 = t_0 - \left(T_0 + nP\right)
\end{equation}
\noindent where $n$ is the number of elapsed transits since the ephemeris transit, $P$ is the period, and $t_0$ is the time of mid-transit for the $n^{th}$ transit. We fit for the offset $\delta t_0$ as opposed to the mid-transit time itself in order to keep the model coefficients within a few orders of magnitude of each other. This is optimal for Levenberg-Marquardt fitting with \texttt{lmfit}.\\
\indent In our full band-integrated white light curve fit from 700 - 1040 nm we see significant features in the residuals. After experimenting with decorrelation parameters and wavelength clipping we conclude that the deep water absorption bands redward of 920 nm are leaving imprints on the white light curve, suggesting changes in precipitable water vapor in Earth's atmosphere during some of our observations. The white light curves presented here do not include these problematic bands and are instead integrated from 700 - 920 nm. \\
\indent At this stage we investigate any transit timing variations by comparing our five derived mid-transit times to those quoted in the discovery paper \citep{Berta-Thompson2015} and subsequent work with MEarth and \textit{Spitzer} \citep{Dittmann2017b} (Figure~\ref{fig:o-c}). The mid-transit times from the \textit{Spitzer} data set reported by \citet{Dittmann2017b} are the BJD$_{\mathrm{-}}$OBS values taken from the \textit{Spitzer} header files. We correct these values to BJD$_{\mathrm{TBD}}$, which accounts for leap seconds. We use the values of $P = 1.6289246$ days and $T_0 = 2457184.55804$ days \citep{Dittmann2017b} to claculate all times of mid-transit.\\
We note that our times of mid-transit are consistently two minutes earlier than the predicted time. We have simultaneous transit observations with the MEarth telescope array, which do not agree with our transit times. This timing offset in this analysis could be due to some unexplored systematic in the instrument or the data reduction. We check our header-time conversions to BJD$_{\mathrm{TDB}}$ multiple times following \citet{Eastman2010}, making sure to account for the exposure and read-out times. As a test we perform a simple data reduction using only polynomials and the \texttt{batman} transit light curve package (i.e., without the \texttt{detrendersaurus} pipeline) and were unable to derive transit times in agreement with those of MEarth and \textit{Spitzer}. \\
\indent This discrepancy does not affect our results with respect to the atmospheric analysis since we fix the times of transit to the best fit values when performing our atmospheric analysis, and the time of mid-transit does not affect the transit depth at the time resolution of our data.\\
\indent We compare our derived values of the planet-to-star radius ratio $R_p/R_*$, period $P$, inclination $i$, and scaled orbital distance $a/R_*$ to those reported by \citet{Dittmann2017b} and find that our results are in agreement (Table~\ref{tab:fitparams}). We present the raw white light curves, jointly-fit white light curve, time-binned white light curve, and time-binned white light curve residuals in Figure~\ref{fig:whitelc}.

\begin{table}
\centering
\caption{White light curve derived transit model values, compared to \citet{Dittmann2017b} \label{tab:fitparams}}
\begin{tabular}{LLL}
\tablewidth{0pt}
\hline
\hline
\mathrm{Parameter} & \mathrm{Value\ (this\ work)} & \mathrm{Value\ (D17a)}\\
\hline
\delta t_{0,1}\ \mathrm{[days]} & -0.0015^{+0.00022}_{-0.00023} & --- \\
\delta t_{0,2}\ \mathrm{[days]} & -0.0017^{+0.00017}_{-0.00017} & --- \\
\delta t_{0,3}\ \mathrm{[days]} & -0.0017^{+0.00055}_{-0.00054} & --- \\
\delta t_{0,4}\ \mathrm{[days]} & -0.0016^{+0.00027}_{-0.00027} & --- \\
\delta t_{0,5}\ \mathrm{[days]} & -0.0017^{+0.00017}_{-0.00018} & --- \\
R_p/R_*  & 0.0488^{+0.0012}_{-0.0009}   &  0.0455^{+0.0006}_{-0.0006} \\
P\ \mathrm{[days]}  & 1.62893^{+0.00013}_{-0.00013} & 1.6289246^{+0.0000024}_{-0.0000030}\\
i\ \mathrm{[deg]}  & 88.54^{+0.90}_{-0.90} & 88.68^{+0.40}_{-0.33}\\
a/R_* & 15.91^{+1.236}_{-1.761} & 16.54^{+0.63}_{-0.71}\\
2u_0 + u_1   & 0.876^{+0.012}_{-0.012}       & --- \\
u_0 - 2u_1  & -0.371^{+0.055}_{-0.056}       & --- \\
s_1 & 4.27^{+0.19}_{-0.18} & --- \\
s_2 & 2.73^{+0.12}_{-0.12} & --- \\
s_3 & 6.08^{+0.28}_{-0.26} & --- \\
s_4 & 5.24^{+0.24}_{-0.22} & --- \\
s_5 & 2.89^{+0.13}_{-0.12} & --- \\
\hline
\end{tabular}
\end{table}

\subsubsection{Wavelength-binned light curves} \label{subsubsec:transmissoin}

We investigate the atmosphere of GJ 1132b by creating a transmission spectrum. We divide our light curves into 20 nm wavelength bins and jointly fit for $R_p/R_*$ and the uncorrelated quadratic limb darkening coefficients $2u_0 + u_1$ and $u_0 - 2u_1$, along with the systematic parameters for each respective data set. We fix the times of mid-transit $t_0$ for each night to the values determined from the white light curve fit. We fix the values of $P$, $i$, and $a/R_*$ in our binned wavelength fits to those reported by \citet{Dittmann2017b} as these are derived from a higher resolution \textit{Spitzer} time-series.\\
\begin{figure}
\includegraphics[scale=.42]{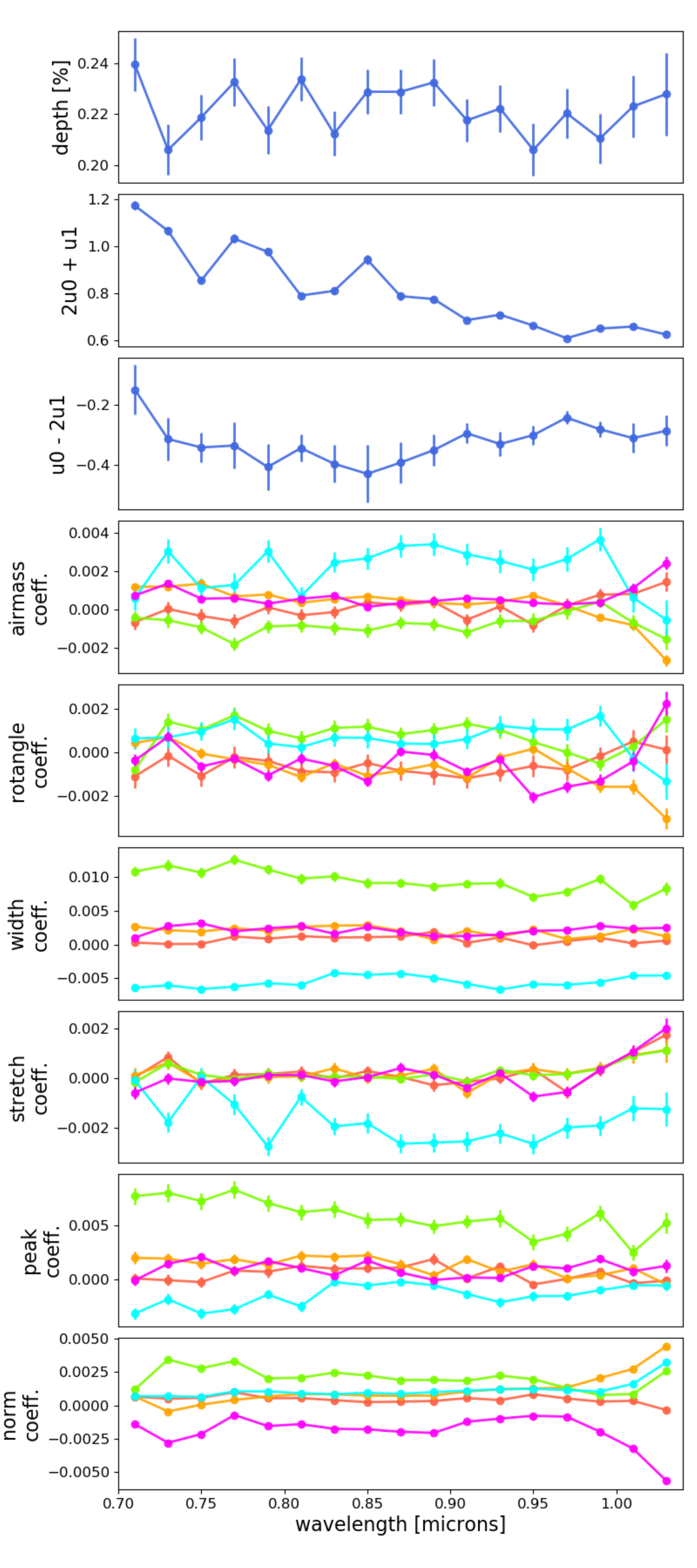}
\caption{Derived parameters in each of the 17 20 nm wavelength bins used in the transmission spectrum. Our joint fit produces a single value for the transit depth and the two uncorrelated quadratic limb-darkening parameters, along with 1$\sigma$ uncertainties, for each wavelength bin (top 3 panels). We also independently fit for the coefficients associated with the decorrelation parameters (Table~\ref{tab:detrendparams}) in each data set in every wavelength bin (bottom 6 panels, colors correspond to those in Figure~\ref{fig:whitelc}). We do not see correlations between the coefficients and the transit depth as a function of wavelength bin.} \label{fig:params}
\end{figure}
\indent Our joint fit produces single values for the radius ratio $R_p/R_*$ and the uncorrelated quadratic limb darkening coefficients $2u_0 + u_1$ and $u_0 - 2u_1$ for each wavelength bin, but each of the five data sets has its own linear fit to the systematics in the light curve. In order to make more meaningful comparisons between the systematic parameters in a given data set we scale each of them by subtracting off the mean value and then dividing by the standard deviation. This ensures that all of our systematic parameters are on the same relative scale and so comparing their fitted coefficients describes the relative importance of each parameter to the fit (Figure~\ref{fig:params}).

\section{Results} \label{sec:results}

\begin{figure*}
\includegraphics[scale=0.42]{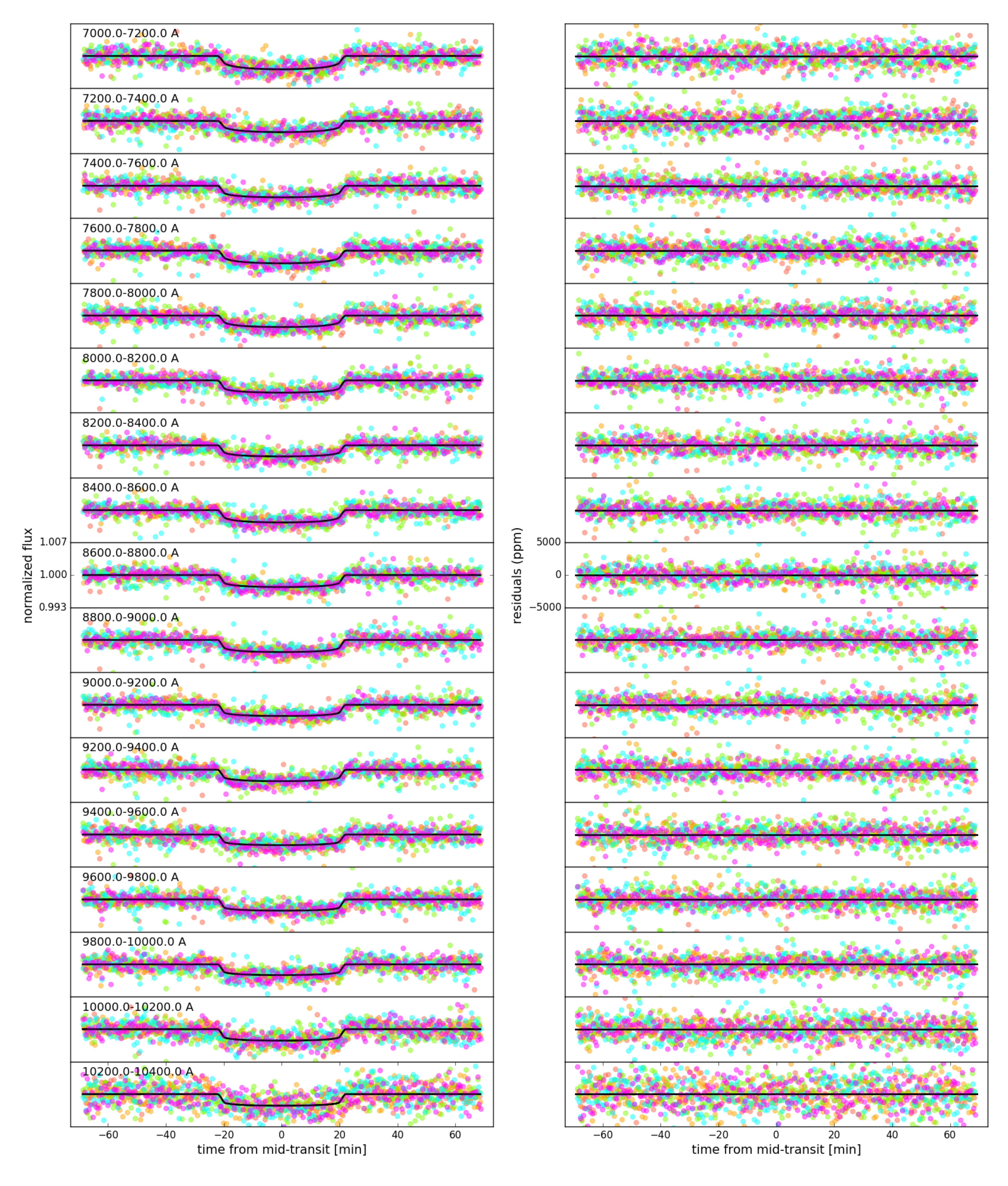}
\caption{\textit{Left}: Detrended light curves (colored points with each color representing one of the 5 data sets used in this analysis) with best fit transit model over-plotted (black lines). The text states the wavelength range in angstroms covered by the light curve directly underneath it. \textit{Right}: Residuals given by the detrended light curves minus the products of the best fit systematics models and transit models. For clarity the y-axis labels in both panels are given only for a single light curve, but all light curves and residuals are plotted on the same scale. For reference, the colors correspond to those in Figure~\ref{fig:whitelc} and the transit depths and RMS values for each wavelength bin are given in Table~\ref{tab:derived}.} \label{fig:lcs}
\end{figure*}

\begin{figure*}
\includegraphics[scale=0.4]{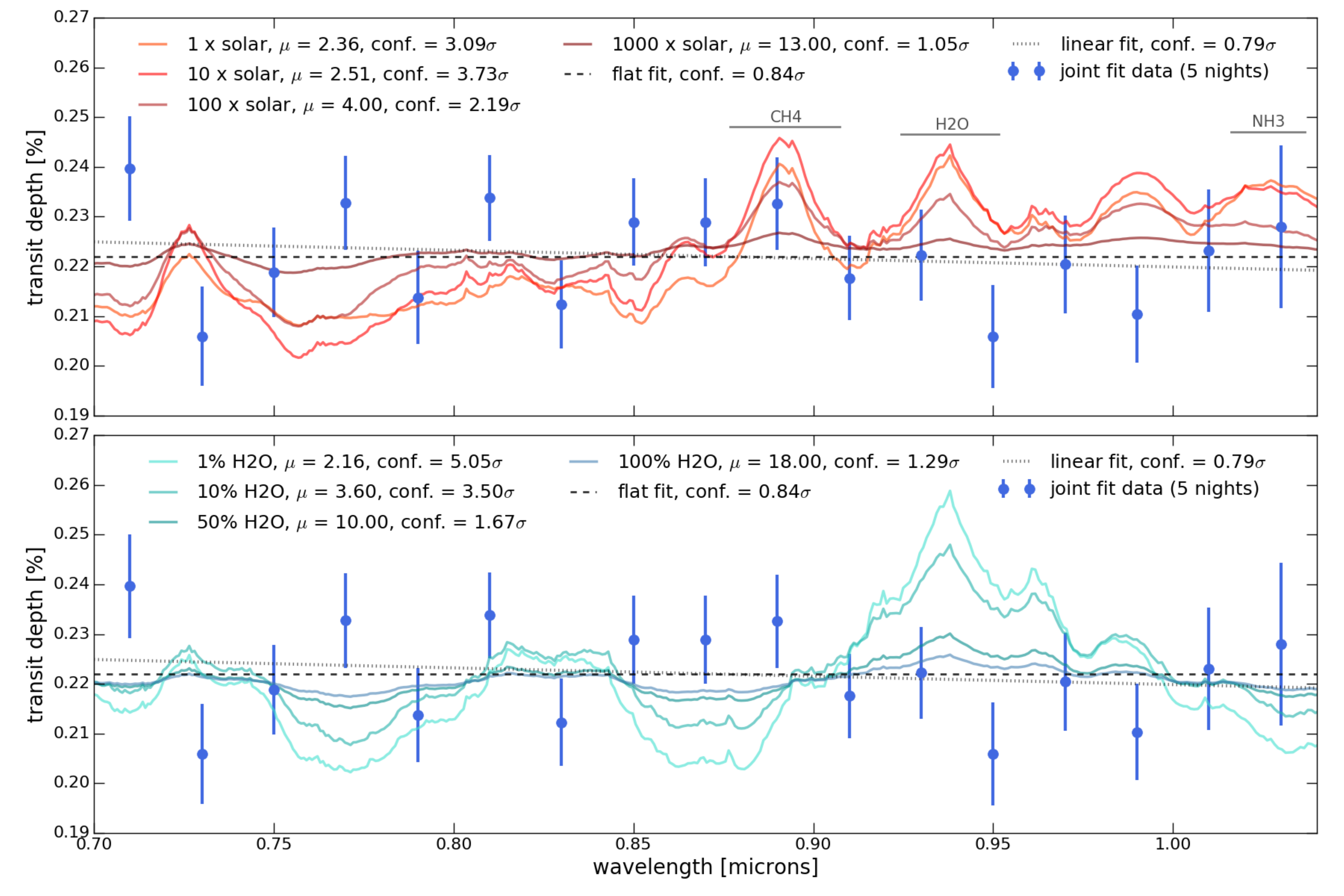}
\caption{Transmission spectrum of GJ 1132b with 1$\sigma$ error bars derived from a joint fit of the five data sets analyzed in this work (both \textit{top} and \textit{bottom}). \textit{Top}: We compare the GJ 1132b transmission spectrum to four clear, solar composition models at 1, 10, 100, and 1000$\times$ solar metallicity by volume. We label the molecular sources of the most prominent features in the model spectra. \textit{Bottom}: We compare the GJ 1132b transmission spectrum to four clear, H$_2$ and H$_2$O models where H$_2$O makes up 1, 10, 50, and 100\% of the atmosphere by volume. All features in these models are due to H$_2$O. Both figures also compare the GJ 1132b transmission spectrum to a flat line at the inverse-variance weighted-average transit depth (black dashed line) and a 1-degree polynomial fit to the transit depths (black dotted line). In the legends of each figure we provide the mean molecular weights of the atmospheres used to create the model transmission spectra and confidences to which the meausred GJ 1132b transmission spectrum disfavors the model atmospheres. The data disfavor low mean molecular weight atmospheres.} \label{fig:trans}
\end{figure*}

\begin{figure*}
\includegraphics[scale=.5]{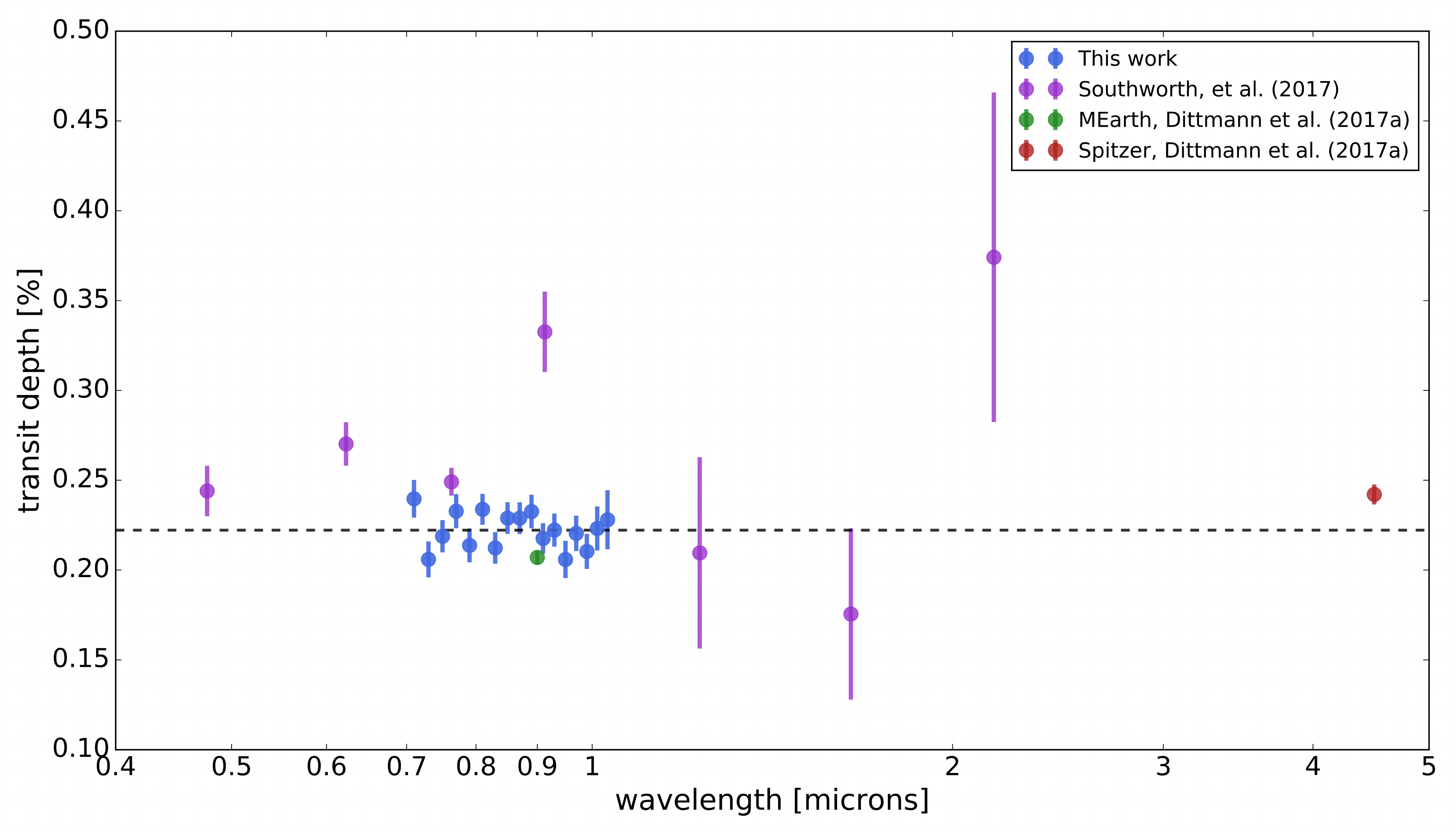}
\caption{The transmission spectrum of GJ 1132b from this work (blue points) with 1$\sigma$ error bars in the context of other GJ 1132b transit data. The dashed line is the inverse-variance weighted average of these transit depths. We plot the photometric transit depths from the MEarth survey (green point) and the \textit{Spitzer} 4.5$\mu$m bandpass (red point) from \citet{Dittmann2017b}, as well as the photometric transit depths in \textit{g}, \textit{r}, \textit{i}, \textit{z}, \textit{J}, \textit{H}, and \textit{K} bands (purple points) from \citet{Southworth2017}. } \label{fig:all}
\end{figure*}

In Figure~\ref{fig:lcs} we present our light curves after dividing out the systematic models for each data set. The wavelength boundaries, RMS, transit depth, and median factor of the expected photon noise limit for each wavelength bin are given in Table~\ref{tab:derived}. According to Figure 2 of \citet{Stefansson2017}, our observations of GJ 1132 are limited by the photon noise so we did not estimate the scintillation noise for the analysis. Including scintillation noise would not change the resulting transit depths but it would decrease our values in the final column of Table~\ref{tab:derived}.\\
\indent Across the 17 wavelength bands we achieve a median transit depth error of 90 ppm. We compare this to 80 ppm for two GJ 1132b transits with the \textit{Spitzer} 4.5$\mu$m channal and 55 ppm with 25 MEarth transits in its photometric band \citep{Dittmann2017b}. \\
\begin{table}
\centering
\caption{Best-Fit Transit Depths}\label{tab:derived}
\begin{tabular}{CCCC}
\tablewidth{0pt}
\hline
\hline
\mathrm{Wavelength} & \mathrm{RMS}       & \mathrm{Transit\ Depth} & \times\ \mathrm{Expected}\\
\mathrm{[\AA]}  & \mathrm{(ppm)}  & \mathrm{\%}   & \mathrm{Noise}\tablenotemark{a} \\
\hline
7000-7200 & 1311 & 0.240 \pm 0.010 & 1.47\\
7200-7400 & 1288 & 0.206 \pm 0.010 & 1.55\\
7400-7600 & 1148 & 0.219 \pm 0.009 & 1.85\\
7600-7800 & 1213 & 0.233 \pm 0.009 & 1.63\\
7800-8000 & 1193 & 0.214 \pm 0.009 & 1.89\\
8000-8200 & 1093 & 0.234 \pm 0.009 & 1.80\\
8200-8400 & 1118 & 0.212 \pm 0.009 & 1.71\\
8400-8600 & 1141 & 0.229 \pm 0.009 & 1.67\\
8600-8800 & 1111 & 0.229 \pm 0.009 & 1.84\\
8800-9000 & 1171 & 0.233 \pm 0.009 & 2.03\\
9000-9200 & 1102 & 0.218 \pm 0.008 & 1.98\\
9200-9400 & 1186 & 0.222 \pm 0.009 & 1.66\\
9400-9600 & 1271 & 0.206 \pm 0.010 & 1.82\\
9600-9800 & 1261 & 0.220 \pm 0.010 & 2.02\\
9800-10000 & 1187 & 0.210 \pm 0.010 & 1.56\\
10000-10200 & 1510 & 0.223 \pm 0.012 & 1.68\\
10200-10400 & 2088 & 0.228 \pm 0.016 & 1.65\\
\hline
\end{tabular}
\begin{minipage}[t]{0.89\linewidth}
\hfill\break
\tablenotetext{a}{Though we are jointly fitting the five data sets we can estimate the expected photon noise limit and resulting RMS for each data set separately. This column represents the median of the five resulting RMS values divided by the expected photon noise for each data set. These values are similar to the average $s$ values that we fit for for each night and for each wavelength bin. It should be noted that we do not include a claculation of the scintillation noise, so these values are conservative.}
\end{minipage}
\end{table}
We present our transmission spectrum in Figure~\ref{fig:trans} and compare it to two sets of four model transmission spectra generated by the \texttt{Exo-Transmit} open source code \citep{Kempton2017}. As inputs we use custom double-grey temperature-pressure profiles and associated equation-of-state files as well as GJ 1132b's surface gravity and radius at 1 bar of atmosphere and GJ 1132's stellar radius \citep{Miller-Ricci2009,Miller-Ricci2010}. The 1 bar planet radius is smaller than the transit radius by an amount that depends on the atmospheric composition, temperature, and gravity. As these values are uncertain we allow the 1 bar planet radius to float in order to achieve the best transmission model fits to our data. Changing the 1 bar planet radius alters the amplitude of the model features as well as the overall depth of the model. The \textit{Spitzer} data from \citet{Dittmann2017b} can resolve the ingress and egress of a transit of GJ 1132b so we adopt the stellar mass and radius quoted in that paper in order to create the temperature-pressure profiles and model transmission spectra.\\
\indent One set of four model transmission spectra assumes solar elemental abundances (dominant in H$_2$ and He) with metallicities that are 1, 10, 100, and 1000$\times$ solar by volume. In these solar composition atmospheres the dominant sources of opacity that contribute to the transmission features are CH$_4$ and H$_2$O, with modest contributions from NH$_3$, H$_2$S, and K. Higher metallicity atmospheres have higher opacities, which strengthen the model features, but also higher mean molecular weights, which dampen the model features. These competing effects are the reason why they highest amplitude features are associated with the 10$\times$ solar metallicity model.\\
\indent The other set of four model transmission spectra assume H$_2$ and H$_2$O atmospheres where H$_2$O makes up 1, 10, 50, and 100\% of the atmosphere by volume. The solar composition models account for collision-induced absorption but the H$_2$/H$_2$O do not. Given how flat the model transmission spectra are this should not impact the results. All models assume a clear atmosphere (i.e., no aerosols).\\
\noindent We also compare the GJ 1132b transmission spectrum to a flat line at the inverse-variance weighted-average transit depth and to a 1-degree polynomial fit to the transit depths. The wavelength bin-averaged values for the \texttt{Exo-Transmit} models are weighted by the recorded counts of a GJ 1132 spectrum across the same wavelength range. By using an observed spectrum of GJ 1132 we account for the difference in relative brightness of GJ 1132 as a function of wavelength, as well as the telluric features imprinted on the spectrum. Because our wavelength bins are so narrow this weighting is virtually indistinguishable from a simple mean across the model wavelength bins. We use the wavelength bin-averaged values of the model transmission spectra to calculate the $\chi^2$ values associated with the model fits to the measured transit depths.\\
\indent Our results disfavor a clear, 1$\times$ solar metallicity atmosphere at 3.09$\sigma$ (99.80\%) and a clear, 10$\times$ solar metallicity atmosphere at 3.7$\sigma$ (99.98\%) confidence. We disfavor a 10\% H$_2$O, 90\% H$_2$ atmosphere at 3.5$\sigma$ (99.95\%) confidence. Our measured transmission spectrum is consistent with a flat line and with metallicities in excess of $\sim$10$\times$ solar or water abundances greater than $\sim$10\%, for aerosol-free atmospheres. \\
\indent We compare our results to those of other groups (Figure~\ref{fig:all}). Our spectrophotometric transit depths are in agreement with photometric transit depths from the MEarth survey and the \textit{Spitzer} 4.5$\mu$m bandpass \citep{Dittmann2017b}, but not in agreement with the photometric transit depths from the GROND multi-band imager \citep{Southworth2017}.\\

%%%%%%%%%%%%%%%%%%%%%%%%%%%%%%%%%%%%%%%%%%%%%%%%%%%%%%%%%%%%%%%%%%%%%%%%%%%%%%%%%%%%%%%%%%%%%%%%%%%%%%%%%%%%%%%%%%%%%%%%%%%%%%%%%%%%%%%
%%%%%%%%%%%%%%%%%%%%%%%%%% Discussion %%%%%%%%%%%%%%%%%%%%%%%%%%%%%%%%%%%%%%%%%%%%%%%%%%%%%%%%%%%%%%%%%%%%%%%%%%%%%%%%%%%%%%%%%%%%%%%%%%%%
%%%%%%%%%%%%%%%%%%%%%%%%%%%%%%%%%%%%%%%%%%%%%%%%%%%%%%%%%%%%%%%%%%%%%%%%%%%%%%%%%%%%%%%%%%%%%%%%%%%%%%%%%%%%%%%%%%%%%%%%%%%%%%%%%%%%%%%

\section{Discussion} \label{sec:disc}

\begin{figure}
\includegraphics[scale=0.425]{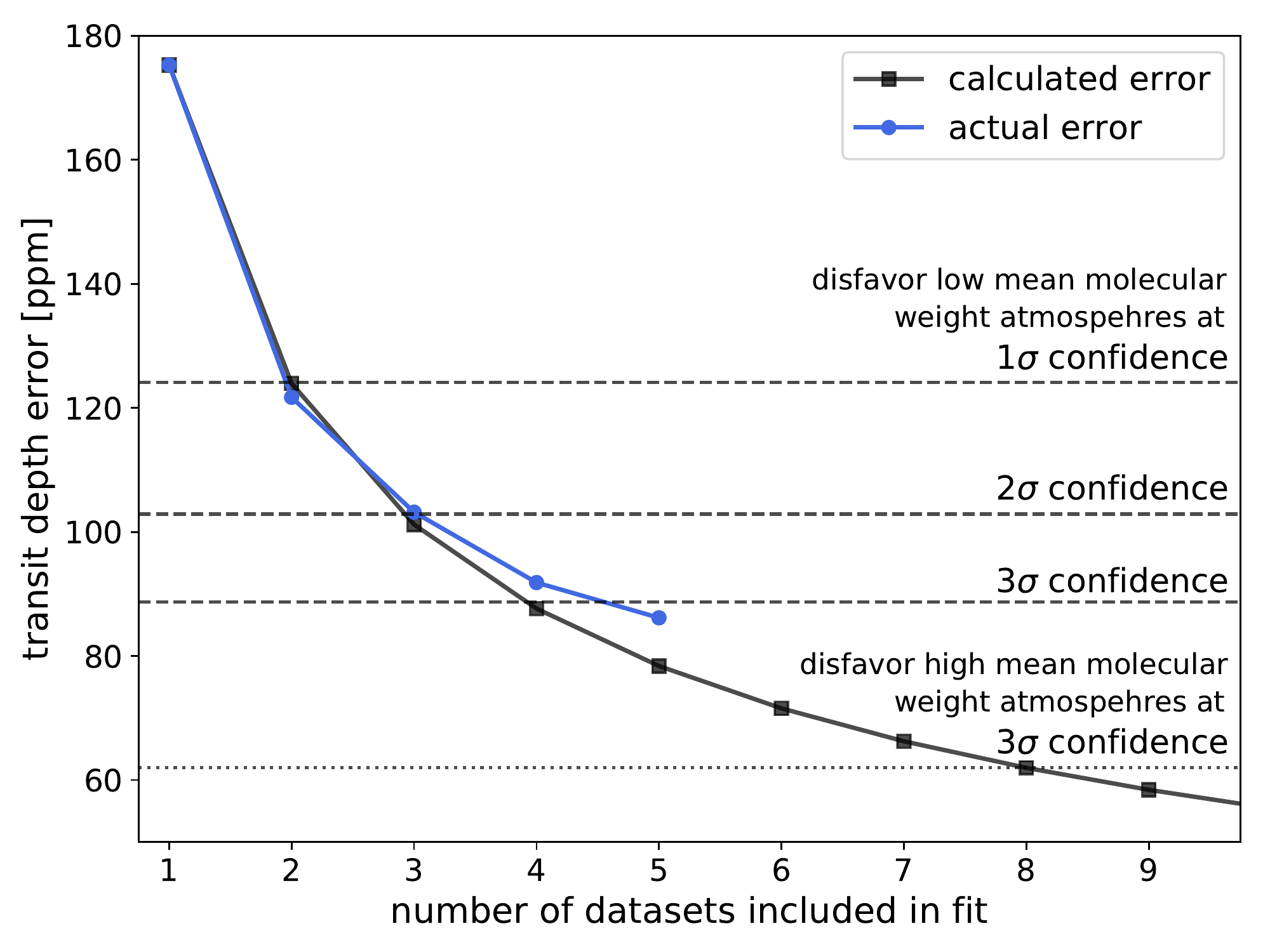}
\caption{Transit depth error as a function of the number of data sets included in the analysis. The blue line and circles shows how the transit depth error decreases when performing the analysis with additional data sets. The black line and squares shows the transit depth error of our first data set divided by the square-root of the number of data sets used in the analysis. We extend the calculated error to investigate what would happen if we captured more than five transits of GJ 1132b. The dashed horizontal lines denote the transit depth error that would disfavor low mean molecular weight atmospheres (10$\times$ solar metallicity and 10\% H$_2$O, 90\% H$_2$) at 1, 2, and 3$\sigma$. We require all five transits to disfavor the low mean molecular weight atmospheres we tested. We theoretically require eight transits to rule out higher mean molecular weight atmospheres (1000$\times$ solar metallicity and 100\% H$_2$O), though likely more given that we do not reach the photon noise limit.} \label{fig:deptherr}
\end{figure}

\subsection{Ground-based detection of terrestrial exoplanet atmospheres}
\indent Our data-reduction process highlights the difficulties of trying to detect terrestrial exoplanet atmospheres from the ground. The signal we are looking for is small (a transit depth of 0.24\% and an atmospheric variation of 0.02\%) and we are not able to reach the photon noise limit (Table~\ref{tab:derived}). One question is whether more data could disfavor higher mean molecular weight atmospheres, or if we needed less data to reach the same conclusions. \\
\indent To answer this question we select a test-case 20 nm wavelength bin, from 830-850 nm, and jointly fit for $R_p/R_*$ and the uncorrelated quadratic limb darkening coefficients $2u_0 + u_1$ and $u_0 - 2u_1$, as we did in our analysis, using 1, 2, 3, 4, and 5 data sets in each fit. We add the data sets in order of decreasing singal-to-noise: first data set 2, then 5, 1, 4, 3. We record the error on the transit depth after each data set is added to the analysis. We compare these to the errors in transit depth from the first data set, scaled by the inverse of the square-root of the number of data sets included.\\
\indent As shown in Figure~\ref{fig:deptherr}, we require all five transits of GJ 1132b to rule out low mean molecular weight atmospheres at high confidence. Theoretically, eight transits are needed to rule out the highest mean molecular weight atmospheres we tested (1000$\times$ solar metallicity and 100\% H$_2$O), though this is a minimum estimate since we do not achieve the photon noise limit and therefore our error bars do not decrease by the square-root of the number of data sets included in the analysis.\\
\indent In the coming era of extremely large ground-based telescopes (ELTs) detecting and characterizing terrestrial exoplanet atmospheres may be in reach. For example, the Giant Magellan Telescope (GMT) will have a diameter of 24.5 m, compared to the 6.5 m diameter of Magellan Clay. This means that the GMT will receive about $(24.5/6.5)^2 = 14.2$ times the number of photons per observation. The science and instrument requirements for the GMT-Consortium Large Earth Finder (G-CLEF), an optical-band echelle spectrograph with a multi-object spectrograph setting and the first-light GMT instrument, suggest that with a single transit observation of GJ 1132b, GMT/G-CLEF would be able to rule out the high mean molecular weight atmospheres we tested in this analysis \citep{Szentgyorgyi2014}. \\
\indent The caveat for all ELT observations is that reaching the photon noise limit from the ground will still be difficult. This difficulty is derived from the need for comparison stars for ground-based transmission spectroscopy. For the nearby systems that we are interested in, there are few comparison stars of similar spectral type and magnitude available. Increasing the field of view of spectrographs would allow for more and possibly better comparison stars, but, for the brightest stars, expanding the field of view sufficiently to include ideal comparisons will likely exceed the design capabilities of spectrographs. For such bright targets it may be worth investing in ground-based high-resolution spectrographs (R $>$ 100,000) which can make atmospheric detections without simultaneous obervations of comparison stars \citep{Snellen2013}.

\subsection{Theoretical atmosphere of GJ 1132b}
\indent It would be surprising if a planet with such a small radius (1.2 $R_{\oplus}$) and high insolation (19$\times$ Earth insolation) possessed a low mean molecular weight atmosphere. Based on thermal evolution models and extreme ultraviolet mass loss, GJ 1132b falls into a class of planets that would be unable to retain a H/He envelope \citep{Lopez2013}. There is statistical evidence from the \textit{Kepler} data set that close-in planets with small radii ($<$ 1.6 $R_{\oplus}$) are rocky and lacking in low-density envelopes \citep{Rogers2015,Fulton2017}.\\
\indent \citet{Schaefer2016} ran models that couple GJ 1132b's atmosphere and interior, allowing for oxygen exchange between the two. They determine that the most likely atmosphere for GJ 1132b is a tenuous one dominated by abiotic molecular oxygen (O$_2$).\\
\indent This arises as follows: water (H$_2$O) in the GJ 1132b atmosphere is photolysed by the intense UV radiation from the GJ 1132 host star. The hydrogen escapes to space, taking some oxygen with it, but the different escape rates along with uptake by the interior mean that some oxygen can combine to form O$_2$ and remain in the planet's atmosphere \citep{Schaefer2016}. Further modeling that includes additional atmospheric gasses such as N$_2$ and CO$_2$ would be of interest.  \\
\indent If the atmosphere of GJ 1132b is dominated by O$_2$ this would be difficult to detect with any currently existing instrumentation. Not only is the mean molecular weight of O$_2$ relatively high ($\mu = 32$) but it also has few spectroscopic features. Fortunately the photolysis of O$_2$ leads to the production of ozone (O$_3$). Given the asymmetry of this molecule it produces higher-amplitude spectroscopic features and is more amenable to detection.
\indent An atmosphere around GJ 1132b may be dominated by other molecules. We see examples in the Solar System of small bodies with high mean-molecular weight atmospheres other than Earth's. Venus, for instance, has a thick atmosphere of CO$_2$ ($\mu = 44$) and Titan has traceable CH$_4$ ($\mu = 16$). These molecules have many prominent spectroscopic features and these atmospheres would be detectable on GJ 1132b in transmission with instruments aboard the \textit{James Webb Space Telescope} (JWST) with 10 transits, according to online predictive tools like \texttt{PandExo} \citep{Batalha2017,Morley2017}. They may also be detectable in transmission with the GMT though the predictive tools are not yet available to test this. Other observing strategies, such as taking emission spectra, will also be useful in constraining the atmospheric properties. \\
\indent With its 19$\times$ Earth insolation and small radius it is likely that GJ 1132b has a high mean molecular weight atmosphere or atmosphere at all. The same can be said for many of the TRAPPIST-1 planets \citep{Gillon2017,deWit2018}. Terrestrial planets farther from their host stars may fare better. LHS 1140b recieves 0.46$\times$ Earth insolation and has a high surface gravity; it therefore may not experience the same rates of atmospheric escape \citep{Dittmann2017a}.

\subsection{Searching for more terrestrial exoplanets}
\indent Perhaps the terrestrial planets with the most accessible atmospheres have not yet been discovered. The GJ 1132, LHS 1140, and TRAPPIST-1 systems are all about 12 parsecs away \citep[respectively]{Berta-Thompson2015,Dittmann2017a,Gillon2017}. \citet{Dressing2015} investigated the occurrence rate of planets around nearby M dwarfs using the full \textit{Kepler} survey and found a cummulative occurance rate of 2.5 $\pm$ 0.2 planets (1 -- 4$R_{\oplus}$) per M dwarf, with periods less than 200 days. So there may be still undiscovered small exoplanets that would be amenable to atmospheric detection and characterization. \\
\indent M dwarfs, with their small sizes, high occurrence rates, and close-in habitable zones, are now the targets of several dedicated transit and radial velocity surveys that aim to identify planets amenable to atmospheric follow-up. Notable transit surveys include MEarth and TRAPPIST \citep{Irwin2015,Gillon2013}, with SPECULOOS and TESS waiting to come online shortly \citep{Burdanov2017,Ricker2015}. Radial velocity surveys focusing on M dwarfs stand to make more detections since they are not as limited by a planet's inclination. Though many of the planets discovered by this method will not transit, their atmospheres may be amenable to phase curve \citep{Koll2016,Kreidberg2016} or high-resolution spectroscopic \citep{Snellen2013} observations. The radial velocity surveys (listed by their acronyms) focused on M dwarfs that are either currently taking data or in the production phase include CARMENES \citep{Quirrenbach2010}, HZPF \citep{Mahadevan2010}, MAROON-X \citep{Seifahrt2016}, NEID \citep{Schwab2016}, NIRPS \citep{Bouchy2017}, and SPIRou \citep{Artigau2014}.\\

%%%%%%%%%%%%%%%%%%%%%%%%%%%%%%%%%%%%%%%%%%%%%%%%%%%%%%%%%%%%%%%%%%%%%%%%%%%%%%%%%%%%%%%%%%%%%%%%%%%%%%%%%%%%%%%%%%%%%%%%%%%%%%%%%%%%%%%
%%%%%%%%%%%%%%%%%%%%%%%%%% Conclusion %%%%%%%%%%%%%%%%%%%%%%%%%%%%%%%%%%%%%%%%%%%%%%%%%%%%%%%%%%%%%%%%%%%%%%%%%%%%%%%%%%%%%%%%%%%%%%%%%%%%
%%%%%%%%%%%%%%%%%%%%%%%%%%%%%%%%%%%%%%%%%%%%%%%%%%%%%%%%%%%%%%%%%%%%%%%%%%%%%%%%%%%%%%%%%%%%%%%%%%%%%%%%%%%%%%%%%%%%%%%%%%%%%%%%%%%%%%%

\section{Conclusion} \label{sec:concl}

\indent We investigate whether or not the small, rocky terrestrial exoplanet GJ 1132b possesses a low mean molecular weight ($\mu \sim$ 2) atmosphere using ground-based telescopes and instrumentation to construct a transmission spectrum. Our analysis disfavors a clear, 10$\times$ solar metallicity and a clear 10\% H$_2$O at high confidence. GJ 1132b likely possesses a high mean molecular weight or depleted atmosphere.\\
\indent While we search for new terrestrial exoplanets we should also continue to learn more about the GJ 1132b atmosphere. Obtaining transits with HST/WFC3 will allow us to confirm the results from this work, especially since space-based telescopes do not have to contend with telluric water features. \citet{Morley2017} suggest that GJ 1132b is the most amenable planet of its kind, currently known, for observation in secondary eclipse with JWST. Small, rocky exoplanets like GJ 1132b challenge our limits of detection and characterization but also present the most exciting opportunities for comparative planetology with the Solar System terrestrial exoplanets, including Earth. \\

%%%%%%%%%%%%%%%%%%%%%%%%%%%%%%%%%%%%%%%%%%%%%%%%%%%%%%%%%%%%%%%%%%%%%%%%%%%%%%%%%%%%%%%%%%%%%%%%%%%%%%%%%%%%%%%%%%%%%%%%%%%%%%%%%%%%%%%
%%%%%%%%%%%%%%%%%%%%%%%%%% Acknowledgements %%%%%%%%%%%%%%%%%%%%%%%%%%%%%%%%%%%%%%%%%%%%%%%%%%%%%%%%%%%%%%%%%%%%%%%%%%%%%%%%%%%%%%%%%
%%%%%%%%%%%%%%%%%%%%%%%%%%%%%%%%%%%%%%%%%%%%%%%%%%%%%%%%%%%%%%%%%%%%%%%%%%%%%%%%%%%%%%%%%%%%%%%%%%%%%%%%%%%%%%%%%%%%%%%%%%%%%%%%%%%%%%%
\noindent This paper includes data gathered with the 6.5 m Magellan Telescopes located at Las Campanas Observatory, Chile. We thank the contributors to the LDSS3C project, the telescope operators and staff at Las Campanas Observatory, and the writers and contributors of the open-source software used in this work. We also thank Mercedes Lopez-Morales, Robin Wordsworth, Dimitar Sasselov, Laura Kreidberg, Kevin Stevenson, and Josh Speagle for helpful comments and conversations. H.D.-L. recognizes support from the National Science Foundation Graduate Research Fellowship Program (grant number DGE1144152). Z.K.B.-T. acknowledges support from the MIT Torres Fellowship for Exoplanetary Research. E.M.-R.K. was supported by the National Science Foundation under CAREER Grant No. 1654295 and by the Research Corporation for Science Advancement through their Cottrell Scholar program. This publication was made possible through the support of a grant from the John Templeton Foundation. The opinions expressed here are those of the authors and do not necessarily reflect the views of the John Templeton Foundation.

\software{lmfit (Newville et al. 2016), dynesty (github.com/joshspeagle/dynesty), batman (Kreidberg 2015), Exo-Transmit (Kempton et al. 2017), mosasaurus (https://github.com/zkbt/mosasaurus), detrendersaurus (https://github.com/hdiamondlowe/detr-endersaurus)}\\

\bibliography{/home/hannah/Harvard-SmithsonianCfA/MasterBibliography}

\end{document}